# Channel matrix, measurement matrix and collapsed matrix in teleportation

XIN-WEI ZHA [*], JIAN-XIA QI and HAI-YANG SONG

School of Science, Xi'an University of Posts and Telecommunications, Xi'an, 710121, P R China

**Abstract** In this paper, two kinds of coefficient matrixes (channel matrix, measurement matrix)associated with the pure state for teleportation are presented , the general relation among channel matrix, measurement matrix and collapsed matrix is obtained. In addition, a criterion for teleportation that the number of coefficient of an unknown state is determined by the rank of the collapsed matrix is given.

## 1. Introduction

Quantum teleportation is a prime example of a quantum information processing task, where an unknown state can be perfectly transported from one place to another by using previously shared entanglement and classical communication between the sender and the receiver. Since the first introduction of quantum teleportation protocol by Bennett [1], research on quantum teleportation has been attracted much attention both in theoretical and experimental aspects in recent years due to its significant applications in quantum calculation and quantum communication. A number of experimental implementations [2-5] of teleportation have been reported and some schemes of quantum teleportation have also been presented [6-9].It is reported teleportation of an arbitrary single qubit, through an entangled channel of Einstein-Podolsky-Rosen (EPR) pair[1]. For a general case, it is proved that only maximally entangled pure states in $C_d \otimes C_d$ could faithfully teleport an arbitrary pure state in $C_d$ [10, 11].

On the other hand, the entanglement is generally considered as a key resource in quantum teleportation. The main thing we need to do this is to know if arbitrary quantum channels can be employed to teleport a multiqubit state. It is shown that three-qubit GHZ state and a class of W states can be used for perfect teleportation of one qubit state [12, 13]. In addition, the tensor products of two Bell states [14] and the genuine four-qubit entangled states [7] are also used for perfect teleportation of two-qubit states. They are all maximally entangled pure states between Alice and Bob. In Ref. [8] we have shown that there exists a relation between the determinant of transformation operator and the SLOCC (stochastic local operations and classical communication) invariant for Bell basis measurement. In Ref.[10], we have shown that there exists a relation between the LU (local unitary) transformation invariant and perfect teleportation. However, in a realistic situation, the teleportation channel is not always the maximally entangled states. Especially, for asymmetric quantum channel, it is important to choose appropriate measurement basis so that the an unknown state can be teleported optimally. Recently, Li et al. [11] have studied SLOCC classification for general n-qubit states via the invariance of the rank of the coefficient matrix. In Ref [12] they provided an efficient method for detecting genuine multipartite correlations of arbitrary n-qubit system in terms of the rank of coefficient matrices. In Ref [15]Wang et al. have studied multipartite entanglement under SLOCC and proposed the entanglement classification under SLOCC for arbitrary-dimensional multipartite (n-qudit) pure states via the rank of coefficient matrix, together with the permutation of qudits. In this paper, we define coefficient matrix of channel matrix, measurement matrix and collapsed matrix in teleportation, furthermore, the general relation among channel matrix, measurement matrix and collapsed matrix is obtained. By calculating the rank of coefficient matrix, we can obtain a criterion for teleportation an unknown state by the rank of the collapsed matrix. We investigate several examples and give optimal match of measuring basis and quantum channel is also to be studied.

---

[*] Corresponding Author. Tel: +86-29-88166094.*E-mail address:* zhxw@xupt.edu.cn (Xin-wei Zha)

## 2. The channel matrix and measurement matrix of teleportation

We first review the construction of coefficient matrix as introduced in Ref.[17,-19]. let us suppose that the sender Alice has particles in an unknown state:

$$|\chi\rangle_a = (x_0|0\rangle + x_1|1\rangle + \cdots + x_{n-1}|n-1\rangle)_a \tag{1}$$

where the coefficient are complex and satisfy normalized condition $\sum_{i=0}^{n-1}|x_i|^2 = 1$.

Suppose that Alice and Bob share a state $|\psi\rangle_{BA}$ as the quantum channel. $|\psi\rangle_{BA}$ can then be represented in the following form:

$$|\psi\rangle_{BA} = \sum_{i,j} C_{i,j} |i\rangle_B \otimes |j\rangle_A, i = 0,1,\cdots,n-1; j = 0,1,\cdots,m-1; \tag{2}$$

For convenience, we write $A_1, \cdots$ as $A$, $B_1, \cdots$ as $B$. That is to say, the particles $A(A_1, \cdots)$ are in Alice's possession, and the other particles $B(B_1, \cdots)$ are in Bob's possession.

A coefficient matrix associated to the state $|\psi\rangle_{BA}$ is defined by

$$C_{n,m}(|\psi\rangle_{BA}) = (C_{i,j})_{n \times m} \tag{3}$$

which can be called channel matrix.

The system state of the particles can be expressed as:

$$|\psi\rangle_s = |\chi\rangle_a \otimes |\psi\rangle_{BA}. \tag{4}$$

It is well known that a quantum state can be transferred perfectly through a swap operator defined by [13]:

$$P_{aB}\langle aB| = \langle Ba| \tag{5}$$

where

$$P_{aB} = \sum_{kl} |kl\rangle_{aB}{}_{aB}\langle lk| \tag{6}$$

Then $|\psi\rangle_s$ can then be represented in the following form [14]:

$$|\psi\rangle_s = |\chi\rangle_a \otimes |\psi\rangle_{BA} = P_{aB} |\chi\rangle_B \otimes |\psi\rangle_{aA} \qquad (7)$$

On the other hand, we have [8]

$$|\psi\rangle_s = |\chi\rangle_a \otimes |\psi\rangle_{BA} = \sum_r |\varphi^r\rangle_{Aa} \sigma_B^r |\chi\rangle_B \qquad (8)$$

Let us assume Alice's measurement basis is $|\varphi^r\rangle_{Aa}$, which has the following form:

$$|\varphi^r\rangle_{Aa} = \sum_{k,l} M^r_{k,l} |k\rangle_A \otimes |l\rangle_a, k=0,1,\cdots,m-1; l=0,1,\cdots,n-1; \qquad (9)$$

A measurement matrix associated to the state $|\varphi^r\rangle_{Aa}$ is defied by

$$M^r_{m,n}\left(|\varphi^r\rangle_{Aa}\right) = \left(M^r_{k,l}\right)_{m \times n} \qquad (10)$$

From Eqs. (4-8), we can obtain collapsed matrix

$$\sigma_B^r = C_{n,m}\left(|\psi\rangle_{BA}\right) \times \left[M^r_{m,n}\left(|\varphi^r\rangle_{Aa}\right)\right]^* \qquad (11)$$

The criterion for teleporting an arbitrary unknown state can be given in terms of the collapsed matrix .

If the collapsed matrix is full rank, i.e. the collapsed matrix is reversible, the unknown particle entangled state can be teleported. Furthermore, If $\qquad \sigma_B^r = kU_B^r$ , $\qquad (12)$ where $k$ is an arbitrary constant and $U_B^r$ is a unitary matrix .

In this case, Bob can determine the state of unknown particles exactly by the inverse of the transformation operator $(U_B^r)^{-1}$. Consequently the unknown particle entangled state is teleported perfectly, and the successful possibilities and the fidelities both reach unity. If the the collapsed matrix is not full rank, i.e. collapsed transformation operator is not reversible, the general unknown entangled state can not be teleported.

### 3. The rank of the collapsed matrix

In terms of channel matrix and measurement matrix, it can be verified that the following result

holds:

(i) For quantum channel $|\psi\rangle_{BA}$, if $n < m$, assume the rank of the channel matrix $C_{n,m}$ and measurement matrix $M_{m,n}$ are $r_c$ and $r_m$ respectively, then we have $r_c \leq n$, $r_m \leq n$. We refer to the rank of the collapsed matrix as $r_{collap}$, Therefore, the rank of the collapsed matrix $\sigma_B^r$ satisfy $r_{collap} \leq n$, in this case, if $r_c = n$, the collapsed matrix is reversible, the unknown particle entangled state can be teleported.

(ii). If $n > m$, similarly, we have $r_c \leq m$, $r_m \leq m$, then the rank of collapsed matrix $\sigma_B^r$ satisfy $r_{collap} \leq m$. Therefore, we know $r_{collap} < n$. in this case, the collapsed matrix is not full rank, the general unknown entangled state can not be teleported.

(iii). if $n = m$, i.e. for symmetric quantum channel, we have $r_c \leq n$, $r_m \leq n$, then the rank of collapsed matrix $\sigma_B^r$ satisfy $r_{collap} = \min(r_c, r_m)$. In this case, only channel matrix and measurement matrix are all full rank, can collapsed matrix become full rank matrix. That is to say, only quantum channel is maximally entangled and Alice's measurement basis is also maximally entangled state. The unknown particle entangled state can be teleported. Obviously, If quantum channel between Alice and Bob is a bipartite product state, i.e., $|\psi\rangle_{BA} = |\varphi\rangle_B \otimes |\varphi\rangle_A$, the rank of channel matrix $C_{n,m}$ must be $r(C_{n,m}) = 1$. Using above definition, it is easy to show that If two n-qubit pure states quantum channel are LU equivalent, i.e. $|\psi\rangle_{BA} = U_B \otimes U_A |\varphi\rangle_{BA}$, then their channel matrices have the same rank.

the two-qubit system. We have

$$|\varphi\rangle_{12} = a_0|00\rangle + a_1|01\rangle + a_2|10\rangle + a_3|11\rangle \tag{13}$$

If the particle 1 is in Alice's possession, and the particle 2 is in Bob's possession. Namely,

$$|\psi\rangle_{AB} = a_0|00\rangle + a_1|01\rangle + a_2|10\rangle + a_3|11\rangle \tag{14}$$

or

$$|\psi\rangle_{BA} = a_0|00\rangle + a_2|01\rangle + a_1|10\rangle + a_3|11\rangle \tag{15}$$

Then the channel matrix associated to the state $|\psi\rangle_{BA}$ is given by

$$C_{2,2} = \begin{pmatrix} a_0 & a_2 \\ a_1 & a_3 \end{pmatrix} \tag{16}$$

Obviously, if the entangled is channel is an Einstein-Podolsky-Rosen (EPR) pair, the channel matrix is full rank. On the other hand, if Alice's measurement basis are Bell states measurement basis, for example,

$$|\varphi^1\rangle_{Aa} = \frac{1}{\sqrt{2}}(|00\rangle + |11\rangle), \quad |\varphi^2\rangle_{Aa} = \frac{1}{\sqrt{2}}(|00\rangle - |11\rangle),$$

$$|\varphi^3\rangle_{Aa} = \frac{1}{\sqrt{2}}(|01\rangle + |10\rangle), \quad |\varphi^4\rangle_{Aa} = \frac{1}{\sqrt{2}}(|01\rangle + |10\rangle)$$

the measurement matrix can be obtain

$$M_{2,2}^1(|\varphi^1\rangle) = \frac{1}{\sqrt{2}}\begin{pmatrix} 1 & 0 \\ 0 & 1 \end{pmatrix}, \quad M_{2,2}^2(|\varphi^2\rangle) = \frac{1}{\sqrt{2}}\begin{pmatrix} 1 & 0 \\ 0 & -1 \end{pmatrix}$$

$$M_{2,2}^3(|\varphi^3\rangle) = \frac{1}{\sqrt{2}}\begin{pmatrix} 0 & 1 \\ 1 & 0 \end{pmatrix}, \quad M_{2,2}^4(|\varphi^4\rangle) = \frac{1}{\sqrt{2}}\begin{pmatrix} 0 & 1 \\ -1 & 0 \end{pmatrix} \tag{17}$$

It is easy to see that the measurement matrix is full rank. That is to say, for symmetric quantum channel, the Bell states measurement basis is optimal measurement basis,

Another example is the three qubits W state[15]

$$|W\rangle_{123} = (\frac{\sqrt{2}}{2}|001\rangle + a_2|010\rangle + a_4|100\rangle)_{123} \tag{18}$$

where $|a_2|^2 + |a_4|^2 = \frac{1}{2}$.

Assume Alice has particles '1' and '2' and Bob has the particle '3'. Then we have channel matrix associated to the state $|W\rangle_{123}$ is given by

$$C_{2,4} = \begin{pmatrix} 0 & a_2 & a_4 & 0 \\ \frac{\sqrt{2}}{2} & 0 & 0 & 0 \end{pmatrix} \qquad (19)$$

If Alice operates a Von Neumann type measurement using the states $\{|\varphi^i\rangle_{12a}, i=1,2,3,4\}$, which are given by

$$|\varphi^1\rangle_{12a} = (\frac{\sqrt{2}}{2}|001\rangle + a_2|010\rangle + a_4|100\rangle)_{12a}, \qquad (20)$$

$$|\varphi^2\rangle_{12a} = (-\frac{\sqrt{2}}{2}|001\rangle + a_2|010\rangle + a_4|100\rangle)_{12a}, \qquad (21)$$

$$|\varphi^3\rangle_{12a} = (\frac{\sqrt{2}}{2}|000\rangle + a_2|011\rangle + a_4|101\rangle)_{12a}, \qquad (22)$$

$$|\varphi^4\rangle_{12a} = (-\frac{\sqrt{2}}{2}|000\rangle + a_2|011\rangle + a_4|101\rangle)_{12a}. \qquad (23)$$

For Von Neumann type measurement basis, the corresponding measurement matrix can be expressed as follow:

$$M^1_{4,2} = \begin{pmatrix} 0 & \frac{\sqrt{2}}{2} \\ a_2 & 0 \\ a_4 & 0 \\ 0 & 0 \end{pmatrix}, \quad M^2_{4,2} = \begin{pmatrix} 0 & -\frac{\sqrt{2}}{2} \\ a_2 & 0 \\ a_4 & 0 \\ 0 & 0 \end{pmatrix},$$

$$M^3_{4,2} = \begin{pmatrix} \frac{\sqrt{2}}{2} & 0 \\ 0 & a_2 \\ 0 & a_4 \\ 0 & 0 \end{pmatrix}, \quad M^4_{4,2} = \begin{pmatrix} -\frac{\sqrt{2}}{2} & 0 \\ 0 & a_2 \\ 0 & a_4 \\ 0 & 0 \end{pmatrix} \qquad (24)$$

Then we have $\quad \sigma^1_B = C_{2,4} \times M^{1*}_{4,2} = \frac{1}{2}\begin{pmatrix} 1 & 0 \\ 0 & 1 \end{pmatrix}$

$$\sigma^2_B = C_{2,4} \times M^{2*}_{4,2} = \frac{1}{2}\begin{pmatrix} 1 & 0 \\ 0 & -1 \end{pmatrix}$$

$$\sigma_B^3 = C_{2,4} \times M_{4,2}^{3*} = \frac{1}{2}\begin{pmatrix} 0 & 1 \\ 1 & 0 \end{pmatrix}$$

$$\sigma_B^4 = C_{2,4} \times M_{4,2}^{4*} = \frac{1}{2}\begin{pmatrix} 0 & 1 \\ -1 & 0 \end{pmatrix} \quad (25)$$

Obviously, collapsed matrix satisfy eq.(12), which equal constant multiple a unitary matrix. Therefore, Bob can determine the unknown particle entangled state exactly by unitary transformation matrix $(U_B^r)^{-1}$.

Next, we give another example of the state, presented by Yeo and Chua[7] which is called the genuine four-qubit entanglement state, namely

$$|YC\rangle_{1234} = \frac{1}{2\sqrt{2}}(|0000\rangle - |0011\rangle - |0101\rangle + |0110\rangle$$
$$+ |1001\rangle + |1010\rangle + |1100\rangle + |1111\rangle)_{1234}$$

If the particle pairs 1, 2 are in Bob's possession, and the other two particles 3, 4 are in Alice's possession. Then the channel matrix associated to the state $|\psi\rangle_{12(34)}$ is given by

$$C_{4,4}(|\psi\rangle_{12(34)}) = \frac{1}{2\sqrt{2}}\begin{pmatrix} 1 & 0 & 0 & -1 \\ 0 & -1 & 1 & 0 \\ 0 & 1 & 1 & 0 \\ 1 & 0 & 0 & 1 \end{pmatrix}$$

Obviously, the channel matrix is the product of constant and unitary matrix. It is clear that the value rank of this channel matrix is $2^{[n/2]}$ [11]. That is to say, the rank of this channel matrix is maximally, therefore, the unknown particle entangled state can be teleported perfectly, and the successful possibilities and the fidelities both reach unity. But if the particle pairs 1, 4 are in Bob's possession, and the other two particles 2, 3 are in Alice's possession. Then the channel matrix associated to the state $|\psi\rangle_{14(23)}$ is given by

$$C'_{4,4}\left(|\psi\rangle_{14(23)}\right) = \frac{1}{2\sqrt{2}}\begin{pmatrix} 1 & 0 & 0 & 1 \\ 0 & -1 & -1 & 0 \\ 0 & 1 & 1 & 0 \\ 1 & 0 & 0 & 1 \end{pmatrix}$$

Obviously, the rank of the coefficient matrix $C'_{4,4}$ is 2, which is smaller than the full rank 4. Therefore, the general unknown two particle entangled state can not be teleported perfectly. However, some special unknown two-parameter state can be teleported, for example, $|\chi\rangle_{a_1 a_2} = (x_0|00\rangle + x_3|11\rangle)_{a_1 a_2}$ state can be teleported. By analyzing above, we know that for the same multi-qubit entangled state, the entanglement property of quantum channel is completely different due to different particle distribution of Alice and Bob. It can be seen that the rank of the channel matrix influences the channel capacity.

## 4. Conclusion

In this paper, we define coefficient matrix of channel matrix, measurement matrix and collapsed matrix, We have proved that the rank of the channel matrix is an important parameter which determine the channel capacity. We investigate several examples of quantum channel. The general relation between channel matrix, measurement matrix and collapsed matrix is obtained. Furthermore, we have shown that quantum channel can be used for faithful teleportation if and only if collapsed matrix is a unitary matrix. The optimal match of measuring basis and quantum channel can be investigated by collapsed matrix. It is known that the number of parameter in an unknown state must be smaller than the rank of the channel matrix.

Acknowledgements

This work was supported by the Natural Science Foundation of Shaanxi Province of China (Grant No. 2013JM1009).